\documentclass[twocolumn]{aastex63}

\usepackage{bm}
\usepackage{mhchem}
\usepackage{hyperref}
\usepackage{amsmath}
\usepackage{amssymb}
\usepackage{graphicx,epstopdf}
\usepackage{epsfig}
\usepackage{color}
\usepackage{colortbl}
\usepackage{xcolor}
\usepackage{aas_macros}
\definecolor{lightgray}{rgb}{0.9,0.9,0.9}
\definecolor{lightergray}{rgb}{0.95,0.95,0.95}
\definecolor{lightblue}{rgb}{0.9,0.9,1}
\definecolor{lightgreen}{rgb}{0.9,1,0.9}
\usepackage{natbib}

\def\gtaprx {\lower .1ex\hbox{\rlap{\raise .6ex\hbox{\hskip .3ex
	{\ifmmode{\scriptscriptstyle >}\else
		{$\scriptscriptstyle >$}\fi}}}
	\kern -.4ex{\ifmmode{\scriptscriptstyle \sim}\else
		{$\scriptscriptstyle\sim$}\fi}}}
\def\ltaprx {\lower .1ex\hbox{\rlap{\raise .6ex\hbox{\hskip .3ex
	{\ifmmode{\scriptscriptstyle <}\else
		{$\scriptscriptstyle <$}\fi}}}
	\kern -.4ex{\ifmmode{\scriptscriptstyle \sim}\else
		{$\scriptscriptstyle\sim$}\fi}}}

\newcommand{\cutt}[1]{\textcolor{blue}{}}

\newcommand{\Ms}{{\ensuremath{M_{\odot} }}}

\newcommand{\C}{{\ensuremath{^{12}\mathrm{C}}}}

\newcommand{\N}{{\ensuremath{^{14} \mathrm{N}}} }

\shorttitle{GS 3073}
\shortauthors{Nandal et al.}

\begin{document}

\title{1000 - 10,000 \Ms\ Primordial Stars Created the Nitrogen Excess in GS 3073 at $z =$ 5.55}

\author{Devesh Nandal}

\affiliation{Department of Astronomy, University of Virginia, Charlottesville, VA 22904, USA}
\affiliation{Center for Astrophysics, Harvard and Smithsonian, 60 Garden St, Cambridge, MA 02138, USA}

\author{Daniel J. Whalen}

\affiliation{Institute of Cosmology and Gravitation, Portsmouth University, Dennis Sciama Building, Portsmouth PO1 3FX}

\author{Muhammad A. Latif}

\affiliation{Physics Department, College of Science, United Arab Emirates University, PO Box 15551, Al-Ain, UAE (latifne@gmail.com)}

\author{Alexander Heger}

\affiliation{School of Physics and Astronomy, Monash University, Vic. 3800, Australia} 

\begin{abstract}

The advent of the {\em James Webb Space Telescope} has revealed a wealth of new galaxies just a few hundred Myr after the Big Bang, a few of which exhibit unusual N/O ratios that are difficult to explain with stellar populations today.  While Wolf-Rayet stars in multiple-burst populations, very massive or rapidly-rotating primordial stars, general relativistic explosions of metal-enriched supermassive stars, or the precursors of globular clusters can in principle account for the nitrogen excess in the galaxies GN-z11 and CEERS 1019, no known stars or supernovae can explain the far higher N/O ratio of 0.46 in GS 3073 at redshift $z =$ 5.55.  Here we show that the extreme nitrogen abundances in GS 3073 can be produced by 1000 - 10,000 \Ms\ primordial (Pop III) stars.  We find that these are the only candidates that can account for its large N/O ratios and its C/O and Ne/O ratios.  GS 3073 is thus the first conclusive evidence in the fossil abundance record of the existence of supermassive Pop III stars at cosmic Dawn.
  
\end{abstract}



\section{Introduction}

Recent surveys with the {\em James Webb Space Telescope} \citep[{\em JWST};][]{ceers1,jades1,uncvr1} have revealed the existence of high-redshift galaxies with large N/O abundance ratios that are difficult to explain with populations of stars that are known today.  Multiple-bursts of massive star formation \citep{kf24,riz25,thesan}, very massive or rapidly-rotating primordial stars \citep{nan24c,tsia24,nan24b}, general relativistic explosions of metal-enriched supermassive stars \citep[SMSs;][]{nag23b}), or the precursors of globular clusters \citep{cam23,mc24} can in principle account for the supersolar nitrogen to oxygen ratios in the galaxies GN-z11 \citep{bun23,sench24} and CEERS 1019 \citep{lar23}.  However, no known population of stars or supernovae can explain the far higher N/O ratio of 0.46 recently found in GS 3073 at redshift $z =$ 5.55 \citep{ji24}.  

Furthermore, it is not enough to account just for N/O ratios in these galaxies, enrichment scenarios must also explain their C/O and Ne/O ratios, which has yet to be done by any galactic chemical evolution model or simulation.  Nevertheless, massive stars are a logical starting point to understand the origin of the excess nitrogen in high-redshift galaxies, as it and Ne are products of late stage nuclear burning and strong convection is needed to cycle them up to the stellar surface to be shed into the surrounding medium.  Here we model N, C, O and Ne production by 1000 - 10,000 \Ms\ Pop III stars to determine if they can produce the nitrogen excess in GS 3073.  Such stars are thought to form in atomically-cooled gas in moderate Lyman-Werner (LW) UV backgrounds \citep[10 - 1000 $J_{21}$, where $J_{21} =$ 10$^{-21}$ erg s$^{-1}$ Hz$^{-1}$ sr$^{-1}$ cm$^{-2}$; e.g.,][]{latif15a}.  These less intense but more common LW fluxes prevented star formation in the gas until it began to be cooled primarily by Ly$\alpha$ emission but also by H$_2$ \citep{prole24}.  Hybrid Ly$\alpha$ / H$_2$ cooling caused collapse rates that were lower than those in purely atomically-cooled gas that formed 10$^5$ \Ms\ stars \citep{pat23a}, instead producing 10$^3$ - 10$^4$ \Ms\ Pop III stars \citep{latif21a}.  SMSs have long been held to be the seeds of the first quasars at $z >$ 6 \citep[e.g.,][]{smidt18,latif22b}.  In Section 2 we describe our SMS evolution models and their elemental abundances in Section 3.  We conclude with a discussion of enrichment scenarios in Section 4. 

\section{Numerical Method}

We modeled the evolution of 1000 - 10,000 \Ms\ Pop III stars in 1000 \Ms\ increments with the Geneva Stellar Evolution Code \citep[GENEC;][]{e08}.  

\subsection{GENEC}

GENEC is a 1D Lagrangian stellar evolution code that uses the Henyey method \citep{hen65} to solve the equations of stellar structure.  Energy generation and nucleosynthesis is calculated with a 38-isotope reaction network and we use the Helmholz equation of state \citep[EOS;][]{ts00}, which includes contributions from degenerate and non-degenerate relativistic and non-relativistic electrons, electron-positron pair production, and radiation.  Internal opacities are derived from the OPAL tables \citep{opal3} and convection is approximated by mixing-length theory. 

We initialize the stars as 2 \Ms\ hydrostatic $n \sim 3/2$ polytropes with uniform entropy profiles. They have a homogeneous chemical composition with $X = 0.7516$, $Y = 0.2484$, and $Z = 0$ \citep{eks12,nan24a} and a deuterium mass fraction $X_{2} = 5 \times 10^{-5}$ \citep{hle18b,nan23}.  They are initially evolved with accretion rates above $2 \times 10^{-2}$ \Ms\ yr$^{-1}$, the critical rate above which the protostars grow as cool, red supergiants \citep{hos12,hos13,nan23,herr23a}.  These rates are later increased as the stars approach the Hayashi limit for the first time to ensure that they reach their target masses by the end of the pre-main sequence or at the onset of core hydrogen burning, after which accretion is terminated.  We assume cold accretion, in which the entropy of the accreting material is matched to that of the surface of the star \citep{hle18b} and is assumed to be radiated away, as in geometrically thin, cold disks \citep{hle13,hle16}. Any excess entropy in the infalling matter is radiated away before reaching the surface of the star \citep{palla92,hos10}. \noindent
For fixed final mass and fixed ejection rule, the integrated ejecta are insensitive to whether pre--MS accretion is variable or constant. By the end of core--He burning the differences are $\leq 10^{-2}$ in mass fraction for N, O, and Ne, and by core--Si burning they are $\leq 10^{-4}$; the only systematic change is the total age.

As Pop III stars form from pristine gas, we exclude line-driven winds from our models \citep{vink01}.  We evolve all ten stars to the end of core oxygen burning and a few to the end of silicon burning.  The stars are partitioned into up to 420 zones in mass and are adaptively rezoned over their evolution.  The stars were also evolved with adaptive time steps, and a moderate convective overshoot of 0.100 in the pre-main sequence and core hydrogen burning stages but not at later stages. Rotation is not included in these models \citep{hle18a}, but see \citet{nan25a} for initial attempts to include rotation in SMS evolution.  We apply the first-order post-Newtonian Tolman-Oppenheimer-Volkoff (TOV) correction to the equation of hydrostatic equilibrium throughout the evolution of the stars to accommodate GR effects by modifying Newton's gravitational constant, $G$ \citep{fuller86,tyr17,hle18b}:
\begin{equation}
G_{\rm rel} = G \left(1 + \frac{P}{\rho c^2} + \frac{2GM_r}{rc^2} + \frac{4 \pi P r^3}{M_rc^2} \right),
\end{equation}
where $P$ is the pressure, $\rho$ the mass density, $c$ the speed of light and $M_r$ is the mass enclosed by radius $r$.  We have since confirmed by linear stability analysis that none of the stars in our study encounter the GR instability at any point in their evolution \citep{saio24}.

In contrast to \citet{nan24c}, we follow each model at least to core--O burning and for a subset to core--Si burning, which fixes the O and Ne stratification outside the CO core and enables robust \emph{Ne/O} predictions used below.

\section{Elemental Yields}


\begin{figure*}
\centering
\includegraphics[width=0.99\textwidth]{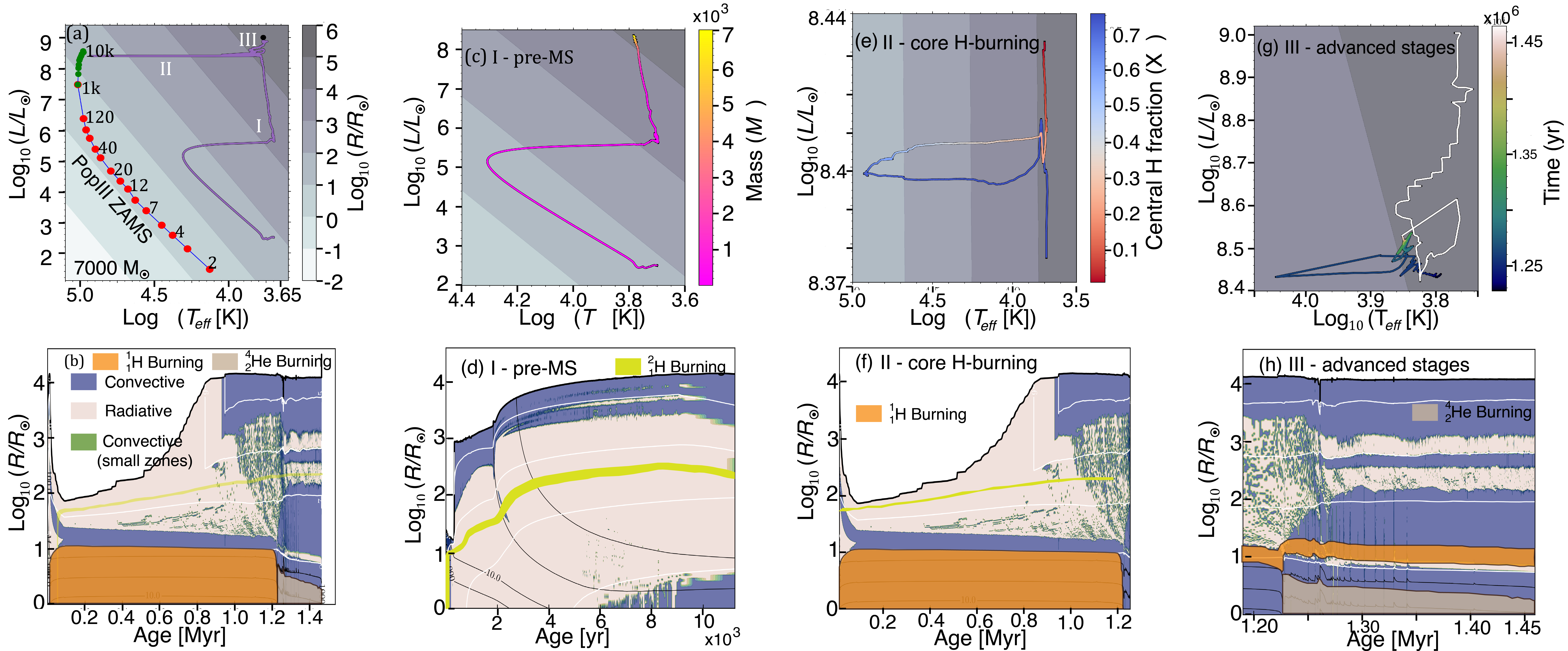} 
\caption{HR (top) and Kippenhahn diagrams (bottom) for the 7000 \Ms\ star.  The entire evolution of the star is shown in (a) and (b), its pre-MS evolution is shown in (c) and (d), its core H burning phase is in (e) and (f), and core He burning to the end of Si burning are in (g) and (h). The black and white lines in the Kippenhahn diagrams are isomass and isothermal lines, respectively, and the red and green dots along the track on the lower left in (a) are the zero-age main sequence (ZAMS) positions of 2 - 120 \Ms\ and 1000 - 10,000 \Ms\ Pop III stars for comparison.}
\label{fig:evol}
\end{figure*}




\begin{figure}
\center
\begin{tabular}{c}
\includegraphics[width=0.45\textwidth]{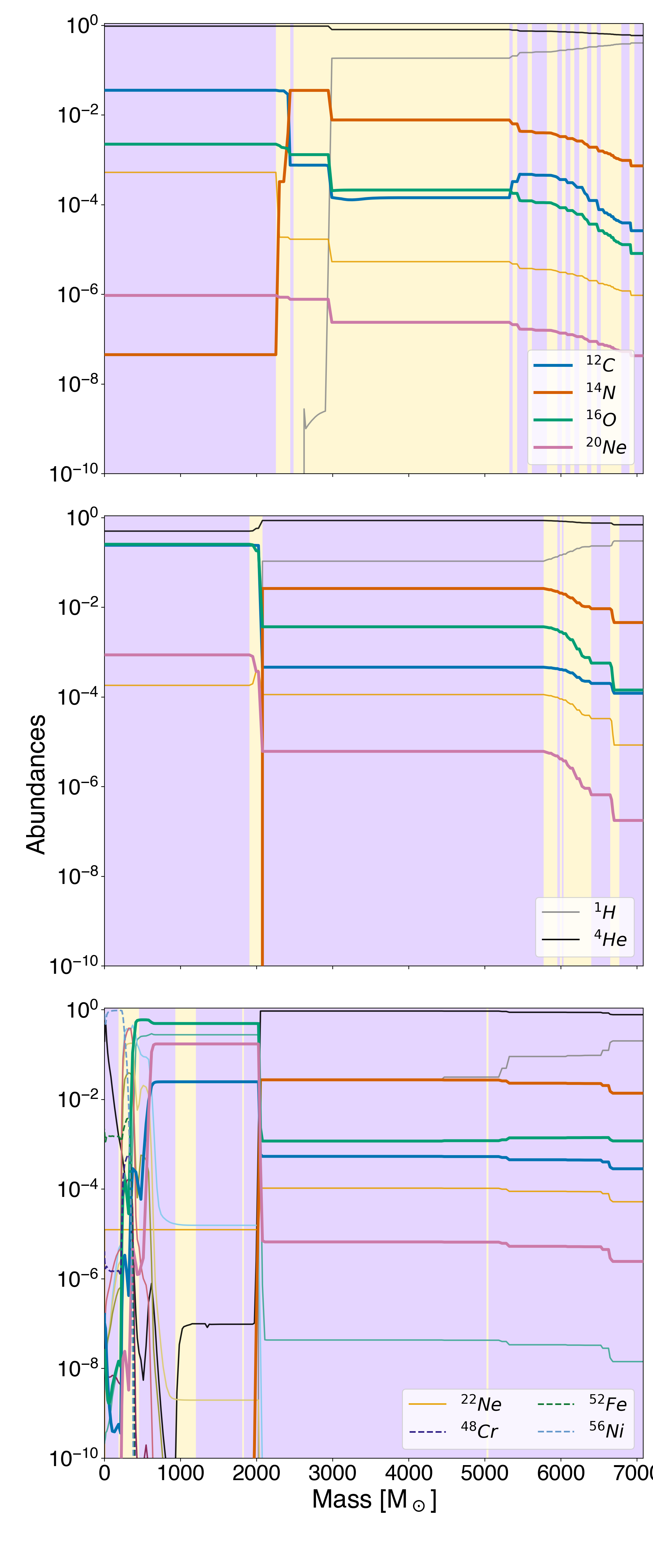}   \\
\end{tabular}
\caption{Abundance profiles of the $7000\,M_\odot$ star at three evolutionary stages. 
\textit{Top}: beginning of core He burning. 
\textit{Middle}: middle of core He burning. 
\textit{Bottom}: end of Si burning. 
Purple and yellow regions mark convective and radiative zones, respectively.}
\label{fig:abun}
\end{figure}

We show HR and Kippenhahn diagrams for the 7000 \Ms\ star as a fiducial case in Figure~\ref{fig:evol} and its abundance profiles during He burning and at the end of Si burning in Figure~\ref{fig:abun}.  The thin radiative layer between the core and the convective H burning shell in panel (h) of Figure~\ref{fig:evol} is permeable to products of nuclear burning, which pass into the H shell and fuse with hydrogen or are dispersed throughout the star.  $^{12}$C from He burning reaches the H shell at 1.22 Myr and fuses into $^{13}$N and then $^{14}$N.  This channel creates the N excess consistent with that observed in GS 3073 because it continues throughout helium burning, the second longest evolutionary stage of the star.  N is soon dispersed throughout the star by the numerous convective zones above the H layer, becoming the third most abundant element at the surface as shown in Figure~\ref{fig:abun}.  $^{16}$O due to alpha captures by $^{12}$C in the core, which can also create $^{14}$N in the H shell, is also mixed throughout the star, becoming the fourth most abundant element at the surface.

$^{14}$N and $^{16}$O synthesis via the CNO cycle depletes $^{12}$C over time.  Although it is the fourth most abundant element at the start of helium burning it falls to fifth place below $^{16}$O halfway through burning as shown in Figure~\ref{fig:abun}.  $^{14}$N produced in the H shell also returns to the convective core and forms $^{22}$Ne after the onset of He burning. This $^{22}$Ne is then mixed throughout the star, becoming the sixth most abundant element at the surface. $^{16}$O in the core from the CNO cycle can capture an alpha and form $^{20}$Ne, but longer $^{16}$O formation times result in less $^{20}$Ne being produced than $^{22}$Ne.  Nuclear ash created in the core after the end of central helium burning never reaches the outer layers of the star because convection timescales are longer than later stages of nuclear burning, which last just a few days.  Heavier elements such as $^{26}$Mg, $^{28}$Si, $^{44}$Ti, $^{48}$Cr and $^{56}$Ni thus remain trapped in the core, as shown in Figure~\ref{fig:abun}.  Final yields for all 10 stars and mass losses (as explained below) are shown in Table 1 in the Appendix.

\subsection{Abundance Ratios / Dilution Factors}

\begin{figure}
\center
\includegraphics[width=0.43\textwidth]{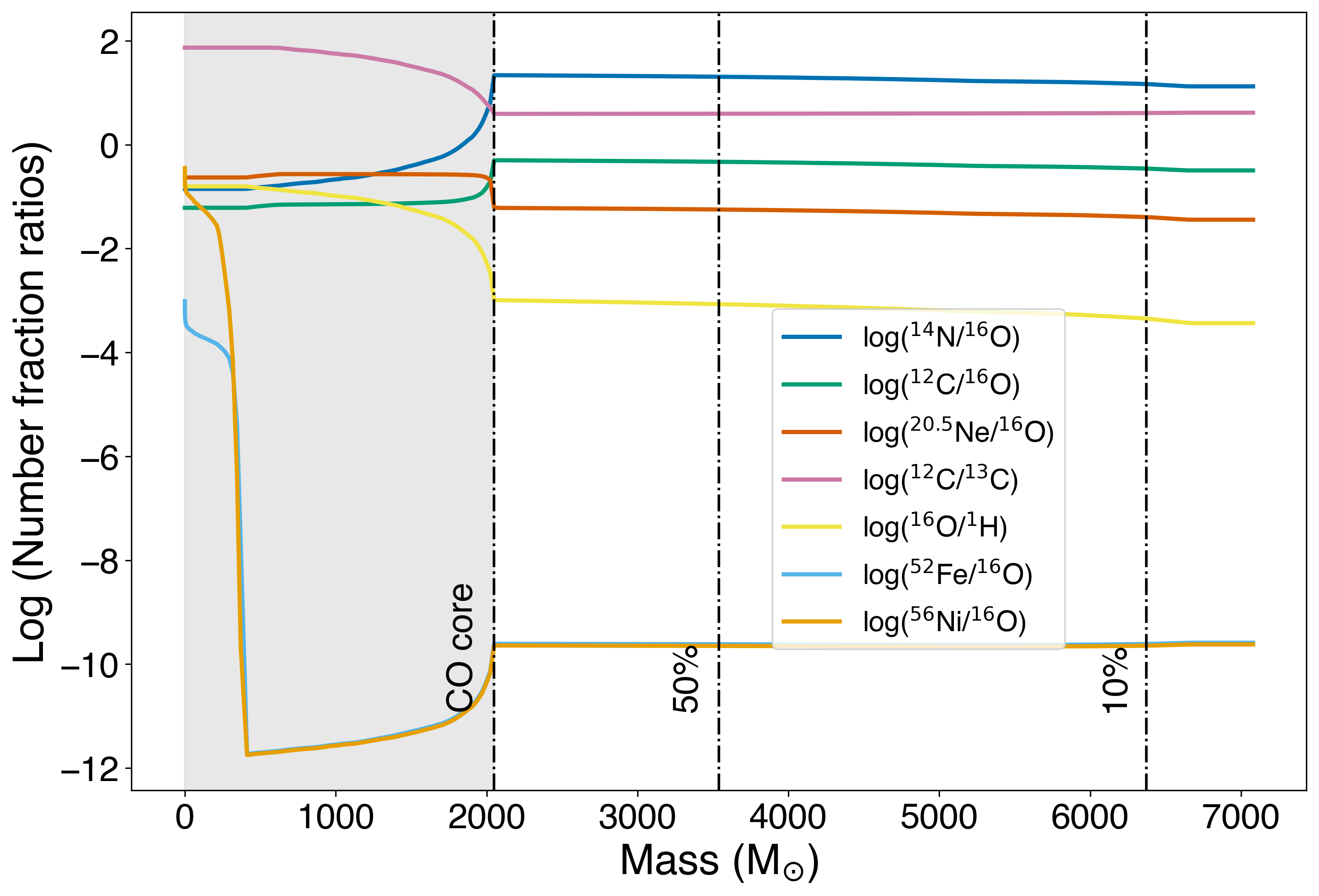} 
\caption{Element number ratios in the star at the end of Si burning.  The masses used to calculate the ratios at a mass coordinate are cumulative, from the stellar atmosphere on the right to the depth corresponding to the given mass loss on the left.}
\label{fig:ratios}
\end{figure}

We calculate N/O, C/O, and Ne/O number ratios for comparison to those observed for GS 3073 with
%
\begin{equation}
\log\left(\frac{\frac{M_X}{A_X}}{\frac{M_Y}{A_Y}}\right) = \log\left(\frac{M_X}{M_Y}\right) + \log\left(\frac{A_Y}{A_X}\right),
\label{eq:ratio}
\end{equation}
where the $M$ and $A$ are the masses (in \Ms) and mass numbers of the elements, respectively. They are plotted versus mass coordinate for the 7000 \Ms\ star at the end of Si burning in Figure~\ref{fig:ratios}, in which the $M$ used in Equation~\ref{eq:ratio} is the total mass of the given element lying above the mass coordinate.  As discussed in the Appendix, mass loss from these stars is likely but not well quantified, so we treat it as a free parameter and consider three fiducial values: 10\% and 50\% of the mass of the star and all the mass above its CO core.  Thus, in Figure~\ref{fig:ratios} N/O $= $ 1.175, C/O $= $ -0.48 and O/H $+$ 12 is 8.65 if 10\% of the mass is ejected.  The number ratios rise as the mass coordinate falls to 2050 \Ms, the CO boundary, after which they diverge in the core because they depend on the evolutionary stage at which the element was created.

\begin{figure}
\center
\begin{tabular}{c}
\includegraphics[scale=0.25]{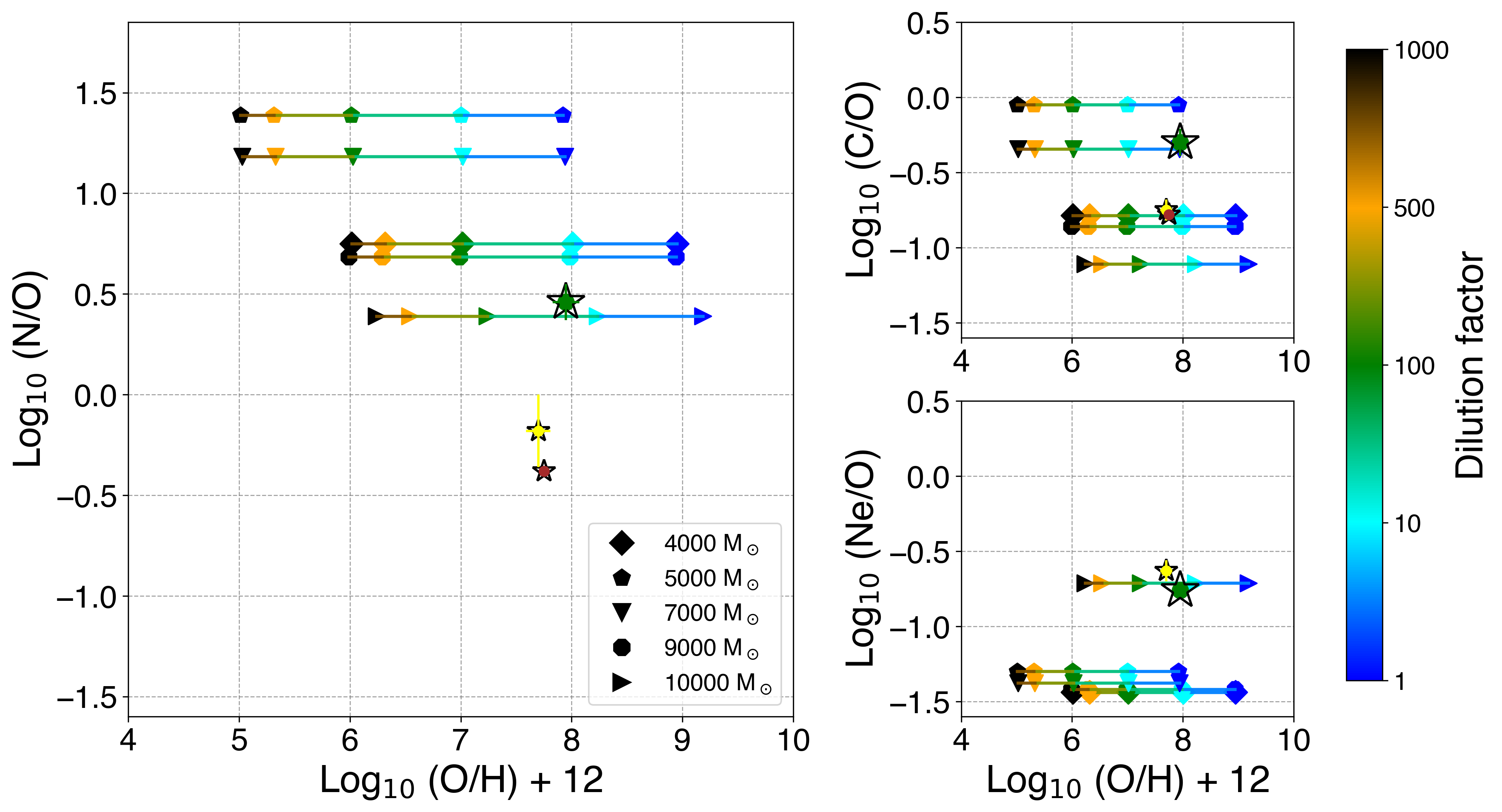} \\
\includegraphics[scale=0.25]{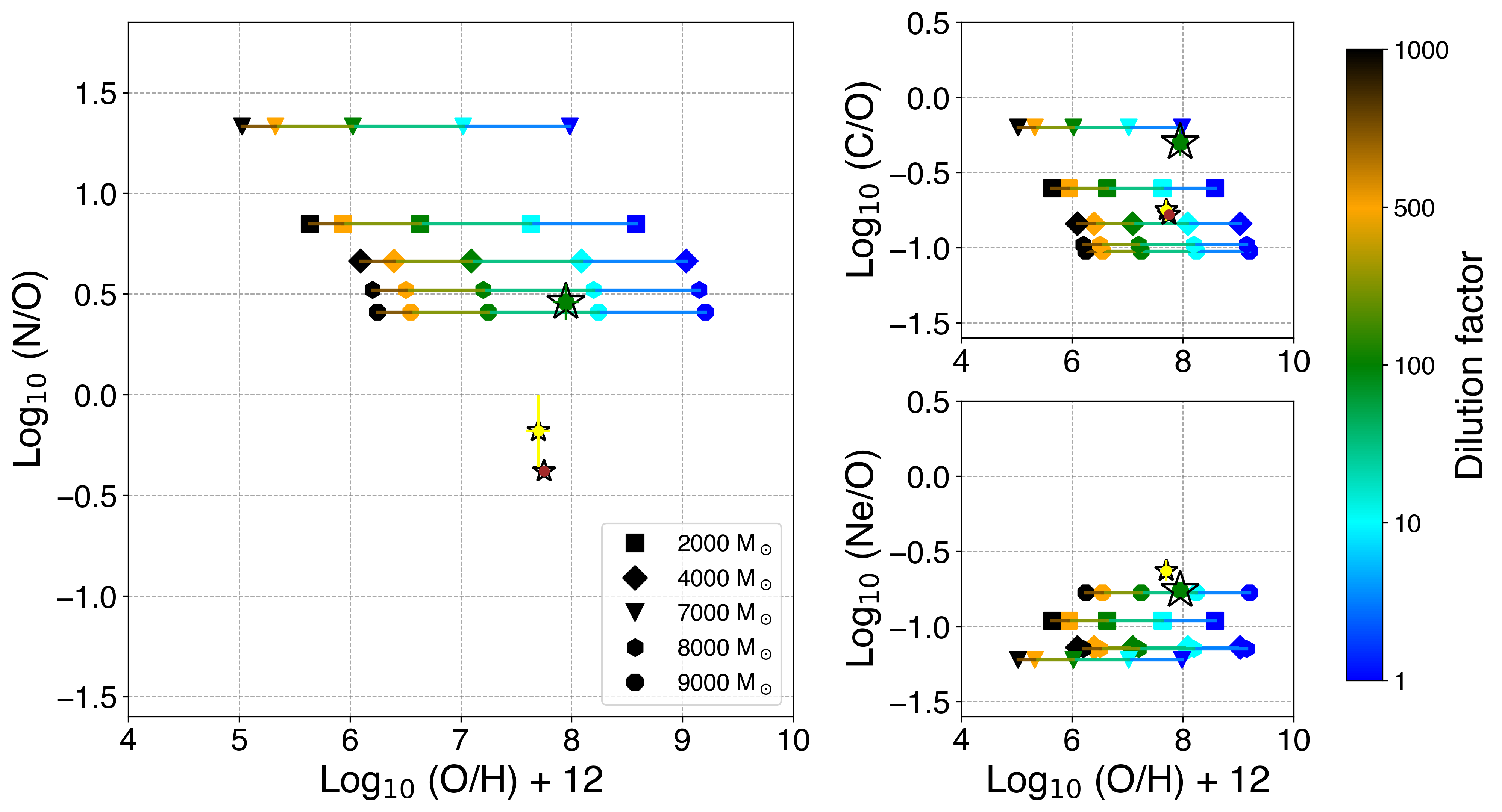} \\
\includegraphics[scale=0.25]{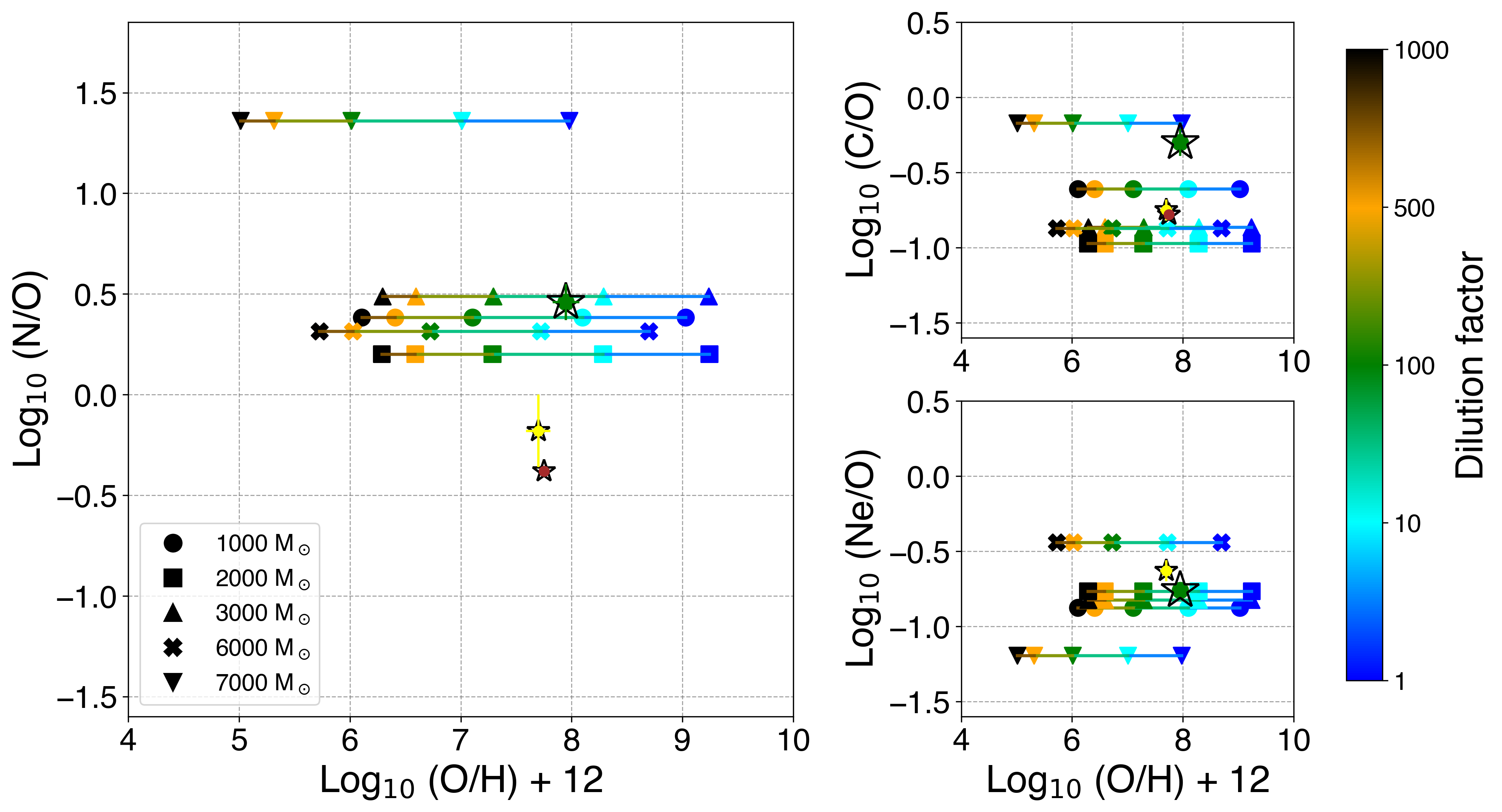} 
\end{tabular}
\caption{N/O, C/O and Ne/O number ratios for 5 select models in the mass range of 1000 - 10,000 \Ms\ Pop III stars.  \textit{Top}: 10\% mass loss.  \textit{Center}:  50\% mass loss.  \textit{Bottom}:  loss of all the mass above the CO core.  The green, yellow and brown stars indicate corresponding ratios for GS 3073, CEERS 1019 and GN-z11, respectively. The dilution factor is the mass of the gas contaminated by outflows from the star divided by the mass of the outflow. Outliers appear only in the ``CO core'' case (bottom panel), where the mass cut can intersect the narrow N rich/O poor layer immediately outside the core; the 10\% and 50\% fixed fraction cases average over this gradient and exhibit much smaller scatter.}
\label{fig:3_10}
\end{figure}

We parametrize enrichment of the galaxy by the star with a dilution factor defined to be the mass of the gas contaminated by ejecta from the star divided by the mass of the ejecta.  We show N/O, C/O and Ne/O ratios measured against the O/H $+$ 12 ratio for all 10 stars as a function of dilution factor from 1 - 1000 for 10\% mass loss, 50\% mass loss, and the loss of all mass above the CO core in the left panels of Figure~\ref{fig:3_10}.  He/H ratios, a prominent observational feature of massive stars, and $^{12}$C/$^{13}$C ratios for all the stars and mass losses are discussed in the Appendix.  The green star marks the observed abundance ratio for GS 3073 derived from rest-frame UV emission lines \citep{ji24}. The yellow and brown stars are corresponding ratios for CEERS 1019 and GN-z11 \citep{mc24,cam23}.  Illumination by a Type-1 active galactic nucleus \citep[AGN;][]{ji24} in GS 3073, and its lower redshift, allow these ratios to be tightly constrained. 

\noindent
Nitrogen enrichment follows a robust sequence in our supermassive models: $^{12}$C produced in the convective He core permeates the thin radiative buffer into the convective H--burning shell, where the CNO cycle converts it to $^{14}$N; intermediate convective zones above the shell then disperse $^{14}$N through the envelope (see Figs.~\ref{fig:evol} and \ref{fig:abun}). The appearance of these zones is controlled by the standard convective instability criterion, $\nabla_{\rm rad}>\nabla_{\rm ad}$, with the radiative gradient
\[
\nabla_{\rm rad} \;=\; \frac{3\,\kappa\,L\,P}{16\,\pi\,a\,c\,G\,m\,T^{4}}\,,
\]
so that near--Eddington luminosities in SMS envelopes favor extended convection without requiring any ad~hoc mixing. After accretion ceases, the structure relaxes on a Kelvin--Helmholtz timescale $\tau_{\rm KH}\!=\!GM^{2}/(RL)$ that is short compared to the duration of core-He burning; as a result the convective topology and the $^{12}$C supply converge across $10^{3}$--$10^{4}$\,M$_\odot$. This produces high N/O with only weak stellar mass dependence in the fixed fraction mass-loss cases (Top and Center panels of Fig.~\ref{fig:3_10}). By contrast, in the ``above CO core'' prescription a small change in the mass cut across the steep N rich/O poor layer just outside the CO core can yield large local excursions in N/O, reflecting a dependence on the composition gradient, rather than on a progenitor mass trend.


\subsection{N/O Ratios}

For mass losses of 10\% our stars produce N/O ratios higher than that of GS 3073 except for 9000 \Ms, which has an N/O ratio of 0.39 at dilution factors of 20 - 30 that falls within the error bars for GS 3073.  The second closest model is the 8000 \Ms\ star, with an N/O ratio of 0.60 at dilution factors of 5 - 15.  For mass losses of 50\%, 8 of the 10 stars produce N/O ratios above those of GS 3073 except for the 9000 and 10,000 \Ms\ models. The N/O ratios of the 8000 and 9000 \Ms\ stars, 0.52 and 0.41, lie within the error bars for GS 3073 at dilution factors of 24 - 46.  If the star sheds all its mass above the CO core only the 3000, 5000 and 7000 \Ms\ stars produce N/O ratios greater than those of GS 3073.  The others still exhibit relatively high N/O ratios, such as 0.1 for the 9000 \Ms\ model. The 1000 and 3000 \Ms\ stars closely match the observed abundances, with N/O ratios of 0.38 and 0.48 for dilution factors of 12 - 38 and 31 - 56, respectively. The extended evolution allows us to match the observed $\log(\mathrm{Ne/O})$ of GS\,3073 while preserving the near--vertical rise in N/O at roughly fixed O/H \citep{ji24}; this diagnostic was unavailable in our earlier work because the models did not reach advanced burning.

\subsection{C/O Ratios}

As shown in the upper right panels of Figure~\ref{fig:3_10}, four stars with 10\% mass loss produce an overabundance of C to O compared to those of GS 3073. The 2000, 5000 and 7000 \Ms\ stars have C/O ratios that are closest to the observed values, -0.45, -0.44 and -0.14, respectively at dilution factors of 5 - 15 at 2000 \Ms\ but only at a dilution factor of 1 for 5000 and 7000 \Ms. For the 50\% case, three models produce an overabundance of C/O with respect to GS 3073. The 7000 \Ms\ model most closely matches GS 3073 with a C/O ratio of -0.30 for dilution factors of 1 - 8. For the loss of all mass above the CO core, only 2 stars produce overabundances of C/O, with the 7000 \Ms\ model again most closely matching GS 3073 with a value of -0.28 over dilution factors of 1 - 8. 

\subsection{Ne/O Ratios}

As shown in the lower right panels of Figure~\ref{fig:3_10}, the 10,000 \Ms\ star with 10\% mass loss is the best match to GS 3073, with Ne/O $=$ -0.71 at a dilution factor of $\sim$ 20.  Only one model here produces an overabundance of Ne/O relative to GS 3073.  For 50\% mass loss the 9000 \Ms\ model is the best match, with Ne/O $=$ -0.77 at a dilution factor of $\sim$ 20. Two of the 10 models produce an overabundance of Ne/O. Finally, for the CO core case two stars match GS 3073, with the 2000 \Ms\ model coming closest with Ne/O $=$ -0.77 at a dilution factor of $\sim$ 20. Four models here produce an overabundance of Ne/O.  
\subsection{Stars Below 1000 \Ms\ or Above 10,000 \Ms}
We find that N/O ratios for GS 3073 cannot be explained by Pop III SMSs above 10,000 \Ms\ or below 1000 \Ms.  900 and 950 \Ms\ stars evolved until the end of core O burning do not produce N/O ratios greater than -0.16, regardless of mass loss, and 15,000 \Ms\ and 21,000 \Ms\ stars evolved to the end of core C and He burning produce an N/O of at most 0.19, still well below the observed 0.46.  The 21,000 \Ms\ star also has a much lower O/H $+$12 ratio of 3 - 6, effectively ruling it out as a candidate. These stars strongly suggest the presence of an upper and lower limit in SMS mass beyond which the GS 3073 N/O ratio cannot be produced. 

\noindent
The emergence of this window is also captured by the integrated ejecta shown in Table~\ref{tab:star_ejecta}. Defining
\[
\begin{aligned}
M_{\rm N}^{\rm ej} &= \int_{m_{\rm cut}}^{M} X_{\rm N}(m)\,\mathrm{d}m,\\
M_{\rm O}^{\rm ej} &= \int_{m_{\rm cut}}^{M} X_{\rm O}(m)\,\mathrm{d}m.
\end{aligned}
\]
with $m_{\rm cut}$ set by the ejection rule, we find that $M_{\rm N}^{\rm ej}$ varies only weakly with $M$ once the post--accretion structure has relaxed, whereas $M_{\rm O}^{\rm ej}$ grows with the size of the convective He core and with the steep $X_{\rm O}(m)$ gradient near the core boundary. The low--mass edge is illustrated by the 1001\,\Ms\ model at 10\% mass-loss, which ejects $^{14}$N$=1.85$\,\Ms, $^{16}$O$=0.11$\,\Ms, but $^{20}$Ne$=0.00$\,\Ms; at 50\% mass-loss $^{20}$Ne remains $0.00$\,\Ms. Thus Ne/O cannot reach the observed value in GS\,3073. At the high--mass edge, the 10024\,\Ms\ model already carries a substantial oxygen load: at 10\% loss $^{16}$O$=20.39$\,\Ms\ out of $M_{\rm ej}=1003.04$\,\Ms\ ($\simeq2.0\%$) with $\log_{10}(\mathrm{Ne/O})\simeq-0.73$, and at 50\% loss $^{16}$O$=158.13$\,\Ms\ out of $M_{\rm ej}=5011$\,\Ms\ ($\simeq3.2\%$) with $\log_{10}(\mathrm{Ne/O})\simeq-0.60$. These oxygen loads drive dilution tracks toward higher O/H and away from the near--vertical locus of GS\,3073 (see Fig.~\ref{fig:enrich}). By contrast, for intermediate masses the nitrogen yield remains high while the oxygen fraction stays modest; for example, at $M=7079$--$9062$\,\Ms\ with 10\% loss we find $^{16}$O/$M_{\rm ej}\simeq0.13$--$1.6\%$. This combination selects $10^{3}$--$10^{4}$\,\Ms\ as the regime that reproduces the observed N/O at nearly fixed O/H without overproducing oxygen or neon, and this conclusion is insensitive to the ejection prescription (10\%, 50\%, or CO--core cut).

\section{Discussion and Conclusion}


\begin{figure}
\centering
\includegraphics[width=0.47\textwidth]{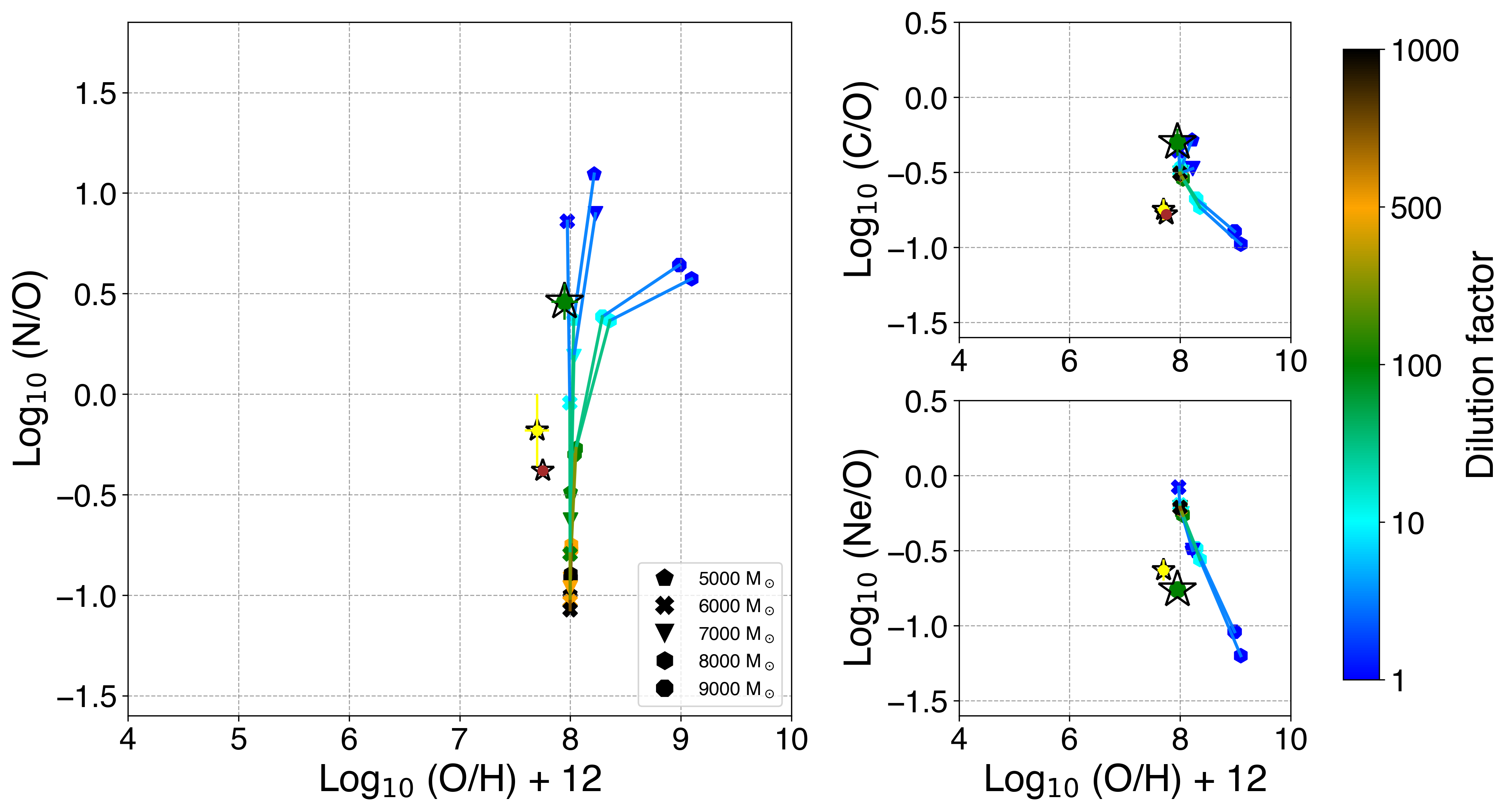} 
\caption{Abundance ratios of 5 SMSs with contributions by the galactic stars for the case of 10\% mass loss. The green, yellow and brown stars indicate corresponding ratios for GS 3073, CEERS 1019 and GN-z11, respectively. If the chemically mature stellar population contributes C, N, O, and Ne mass fractions $\chi_{\rm C} = $ $2.8 \times 10^{-4}$, $\chi_{\rm N} =$ $8 \times 10^{-5}$, $\chi_{\rm O} =$ $1.2 \times 10^{-3}$ and $\chi_{\rm Ne} =$ $9.4 \times 10^{-4}$ then all 10 models can reproduce the observed near vertical transition in N/O ratio while also explaining the C/O ratios \citep{ji24}.}
\label{fig:enrich}
\end{figure}

No single SMS would be expected to match all the abundance ratios of GS 3073 because other generations of stars also contributed to them.  Their contribution decreased N/O ratios and elevated C/O ratios over time because mass loss and SNe produced much more C and O than N \citep{eks12,murphy21}.  SMSs with N/O ratios greater than those in GS 3073 and C/O ratios less than those in GS 3073 can thus still be the origin of its N/O excess and C/O fractions.  We can estimate the contribution to N/O, C/O and Ne/O ratios in GS 3073 from its less massive stars by comparison to surface abundances of halo and disk stars in the MW at similar metallicities \citep{pran18}.  From Figures 4 and 8 of \citet{ji24} we find N and O mass fractions for this population to be 8.0e-5 and 1.2e-3, respectively.  Assuming [Fe/H] $\sim$ -0.6 in GS 3073 \citep{ji24}, we then obtain C/Fe and Ne/Fe ratios and then C and Ne abundances of 2.8e-4 and 9.4e-4, respectively.  We then mix ejecta from the SMS with the ISM for a range of dilution factors. The present grid differs from \citet{nan24c} by advancing to core--O (and in some cases core--Si) burning with TOV corrections, by adding \emph{Ne/O} as a key abundance test, by confronting the tighter abundance set of GS\,3073, and by including an exploratory rotating SMS subset to assess sensitivity to rotational mixing during He burning. Additionally, we also explain the black hole mass of GS3073 by invoking sub-Eddington accretion onto the seed formed by such SMSs at redshifts up to 20.

As shown in Figure~\ref{fig:enrich}, when the SMS and stellar contributions to N/O, C/O, and Ne/O are taken together, we obtain good agreement with those of GS 3073 with a 6000 \Ms\ star with 10\% mass loss in this scenario.  Mixing of the SMS and ISM components produces a nearly vertical shift in N/O versus O/H from its floor of -1.1 to 0.46, suggesting the significant increase in N abundance without a substantial change in O abundance shown in Figure 5 of \citet{ji24}.  This test also shows that the SMSs must be Pop III because no other stars would keep O/H constant.  Ours are also the only known stars that can produce the Ne/O ratio in GS 3073.  While GR supernovae of metal-enriched SMSs can explain the smaller N excess in GN-z11 \citep{nag23b}, they cannot account for its Ne abundances.  GR SNe also blow most of the gas from their host galaxies, greatly delaying subsequent star formation \citep{wet13b,wet13a,wet12d,jet13a,chen14b}, so it is not clear if GS 3073 could have reached its stellar mass, 10$^{9.5}$ \Ms, by $z =$ 5.55 if a GR SN had occurred early in its history.  Furthermore, such an event cannot explain the presence of a BH in GS 3073 because it would not have created one, whereas an SMS can account for both the nitrogen excess in GS 3073 and its AGN.  Not only do several of our stars match Ne/O ratios for GS 3073, some produce an overabundance of Ne that would be diluted down to the observed values by other populations of stars. 

Gas at lower densities in GS 3073, as traced by optical lines, has a much lower N/O $=$ -1.1 than gas at high densities, which is traced by UV lines \citep{ji24}.  Stratified N/O ratios are consistent with our SMS scenario, as cosmological simulations indicate that SMSs form in dense gas in high-redshift halos.  Whether the SMS formed in a satellite halo and later merged with GS 3073 in which a mature population of stars had already formed \citep{regan17a} or it formed in a pristine pocket of gas in GS 3073 \citep{lb20,ven23}, there would have been the near vertical rise in N/O versus O/H.  Because the true distribution of low-metallicity and high-metallicity gas in GS 3073 is not known, it is not yet possible to determine precisely which enrichment scenario happened, only that massive nitrogen production by SMSs must have occurred.

The unprecedented supersolar N/O ratio and C/O and Ne/O ratios in GS 3073 are the first conclusive evidence in the fossil abundance record of the existence of supermassive Pop III stars at the end of the cosmic Dark Ages.  The N/O overabundances of some of our stars also predict that galaxies with even higher N excesses could exist at high redshifts if they are at earlier stages of star formation.  These objects could be found in future surveys by the James Webb Space Telescope and the next generation of ground-based observatories in the decade to come.

\acknowledgments
The authors would like to thank the referee for their invaluable comments and feedback on the manuscript. DN was supported by the Swiss National Science Fund (SNSF) Postdoctoral Fellowship, grant number: P500-2235464 and by the Virginia Initiative of Theoretical Astronomy (VITA) Fellowship. AH acknowledges support by the Australian Research Council (ARC) through grants DP240101786 and DP240103174.  MAL was funded by UAEU via UPAR grant No. 12S111.
 
\bibliographystyle{apj}
\bibliography{refs}

\appendix

\subsection{Protostar Growth} 

The seed of the 7000 \Ms\ star is initialized with accretion rates of 0.025 - 3 \Ms\ yr$^{-1}$, so it evolves as a red supergiant protostar \citep{nan23,herr23a}.  As it grows in mass its central temperatures reach 10$^6$ K and trigger deuterium burning at 3.2 \Ms\ as shown by the yellow band in panel (d) of Figure~\ref{fig:evol}.  Deuterium fusion does not produce enough energy to halt internal contraction so central temperatures continue to rise.  The luminosity and surface temperature of the protostar increase as it grows in mass so its HR track migrates up and to the left in panel (c) of Figure~\ref{fig:evol}.  Rising core temperatures then drive an outward luminosity wave \citep{stahler86b,sak15,nan23} lasting for $\sim$ 100 yr that causes the protostar to expand and cool at 18 \Ms, and the HR track veers back to the right and then evolves along the nearly vertical Hayashi limit.  As shown in panel (d), the protostar has a large convective envelope over a radiative zone for most of its evolution until a central convective zone appears as the star approaches core H burning.

\subsection{Central H Burning}

After accretion is halted at 12,000 yr when the protostar reaches 7000 \Ms, its structure settles on a Kelvin-Helmholtz (KH) timescale $\tau_{KH} = GM^{2}/2RL \sim$ 80,000 yr and it evolves to the left in the HR diagram in panel (e) as it becomes hotter and brighter.  Rising core temperatures due to KH contraction activate central H burning at log$_{10}(T_{\text{eff}} \, \text{[K]}) = 4.3$ and then the triple alpha chain.  Carbon from the triple alpha cycle activates CNO burning, which finally halts contraction and the star settles onto the main sequence.  The model reaches the leftmost point in the HR diagram in (e) at 95,000 yr but then migrates back to the right as energy released by central H burning expands and cools the star.  As shown in panel (f), by 0.1 Myr the star has a structure similar to those of 9 - 120 \Ms\ stars, with a convective core that recedes in mass as central H burning proceeds and a radiative envelope that grows in radius over time.  Changes in adiabatic gradients ($\nabla_{\rm rad}$ and $\nabla_{\rm ad}$) promote the formation of a large number of small convective zones in the radiative region after 0.4 Myr.  A large convective zone also forms in the outer envelope when the star reaches radii log$_{10} (R/R_{\odot}) > 4$. At 1.22 Myr the model has depleted all of its core hydrogen and exits the main sequence.

\subsection{Post-Main Sequence Evolution}

Central helium burning begins at 1.2 Myr at log$_{10} (T_{\text{eff}} \, \text{[K]}) = 3.85$ and log$_{10} (L/L_{\odot}) = 8.50$ in panel (g) of Figure~\ref{fig:evol}. At 1.227 Myr the convective core has a helium-burning center with an outer convective boundary as shown in panel (h). Above the convective core lies a thin radiative zone and the convective H burning shell. Several small convective regions appear in the radiative layer over the course of central He burning.  Above the H burning shell there are also large convective zones, radiative zones with convective cells, and finally very large convective envelopes.  The structure of the star is thus primarily convective at late times.  

Carbon burning begins at 1.35 Myr, before the end of central He burning at 1.465 Myr because high central temperatures allow C to fuse as soon as it is produced by He.  Ne burning, mostly $^{20}$Ne, then begins but only lasts for 10 hours and produces $^{16}$O via photodisintegration at temperatures of $12 \times 10^{9}$ K. Alpha particles from photodisintegration react with $^{20}$Ne to form $^{24}$Mg. Because of its short duration, Ne burning begins and ends at the same point in the HR diagram, at log$_{10} (T_{\text{eff}} \, \text{[K]}) = 3.84$ and log$_{10} (L/L_{\odot}) = 8.51$ in panel (g).  Core O burning then begins and finishes in 41 minutes, with subsequent Si burning that lasts for just a few minutes after which the run is ended at 1.46 Myr.   

\subsection{Collapse}


\begin{figure}
\center
\begin{tabular}{cc}
\includegraphics[width=0.40\textwidth]{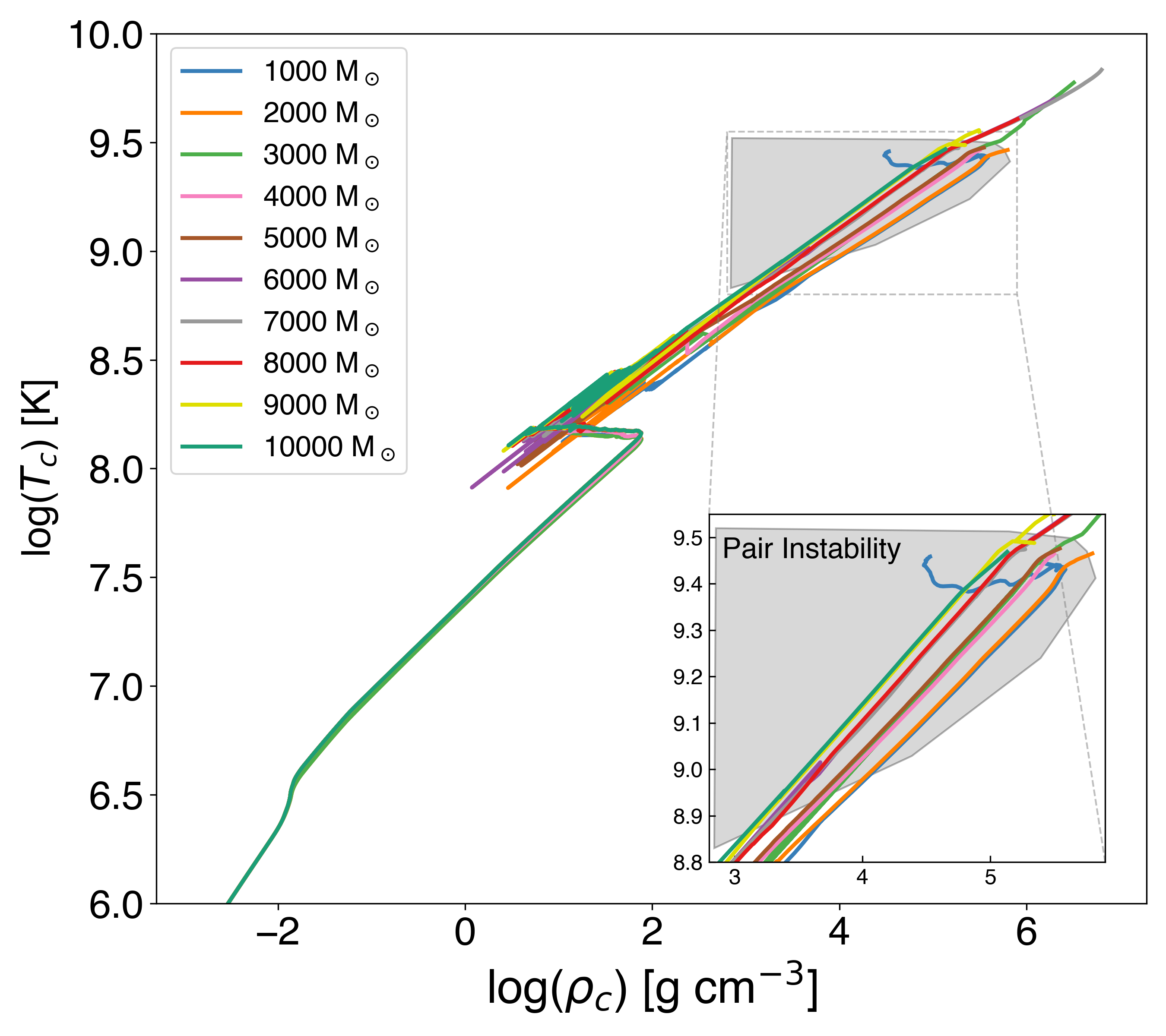} & 
\includegraphics[width=0.48\textwidth]{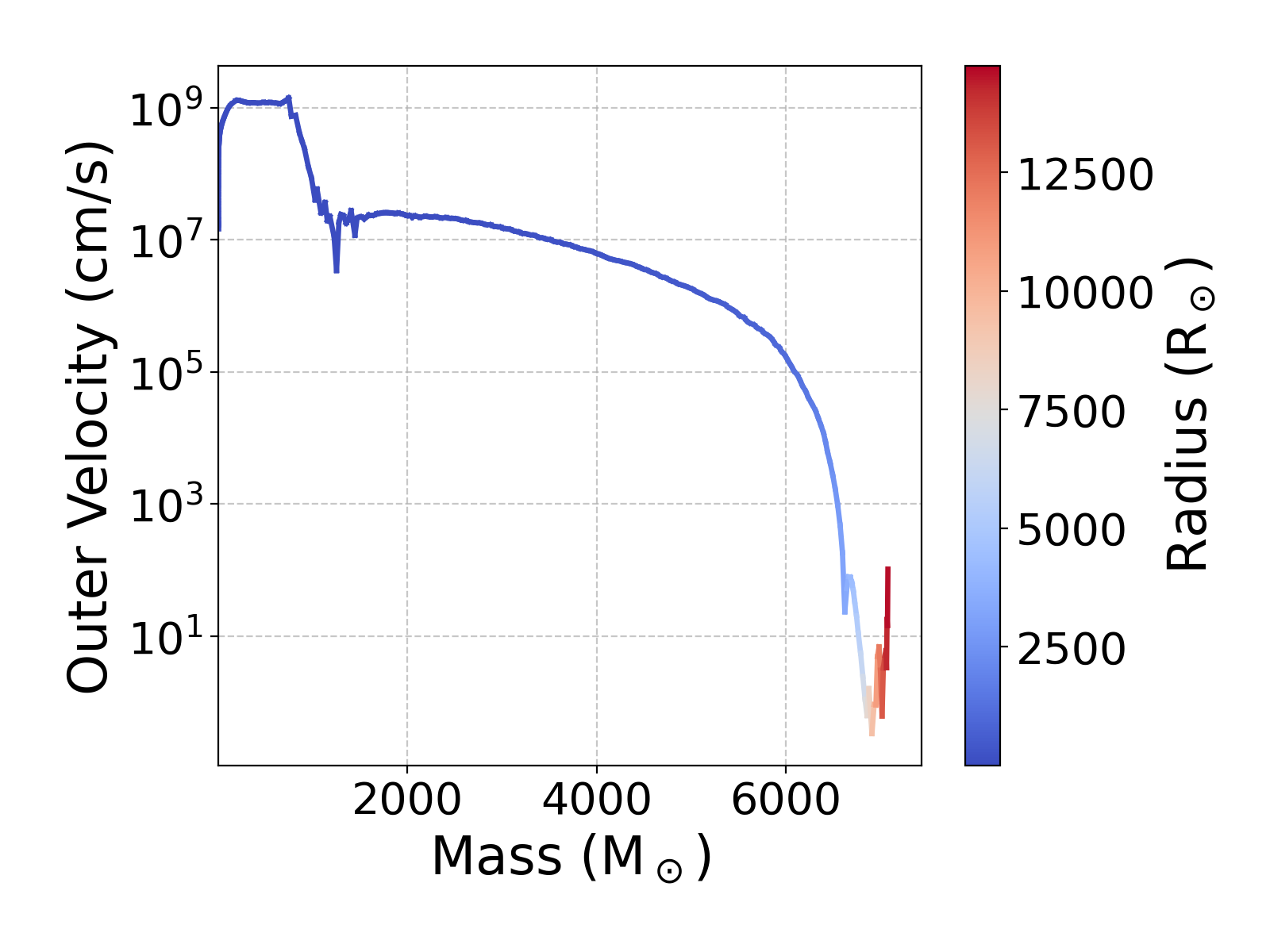} 
\end{tabular}
\caption{\textit{Left}:  evolution of central densities versus central temperatures.  The grey shaded zone is the PI regime.  A few models such as the 6000, 7000, 8000 \Ms\ stars were evolved past the PI into core Si burning.  \textit{Right}:  Collapse of the 8000 \Ms\ star to a BH.}
\label{fig:PI}
\end{figure}

As shown in the right panel of Figure~\ref{fig:PI}, all ten stars enter the pair instability (PI) regime at the end of their lives.  In less massive Pop III stars the PI can lead to violent pulsational mass loss \citep{wbh07,wet13d,chen14a,w17,lnb19,chen23} or an explosion that completely unbinds the star \citep{rs67,brk67,wet12b,wet13e,xing23}.  However, even if the PI triggers explosive thermonuclear burning in O and Si in these stars it is unlikely that the energy released would reverse their collapse because of their large masses.  To test this point we evolved the 8000 \Ms\ star from the end of Ne burning to collapse in the Kepler stellar evolution code \citep{kep1,kep2} because it can transition to hydrodynamics to follow collapse and explosions.  We find that even though the star encounters the PI, photodisintegrations also crack heavy elements into alpha particles and alphas into nucleons in the core.  The corresponding loss of thermal pressure triggers rapid collapse of the core, and the central 1200 \Ms\ of the star reaches infall velocities of 10\% of the speed of light by the end of the simulation, as shown in the left panel of Figure~\ref{fig:PI}.  The kinetic energy of this mass is several MeV/nucleon, more than nuclear burning can yield.  The other stars would likely have similar fates.  Some elements may be explosively synthesized during collapse but they would almost certainly be swallowed by the BH because there is so much mass in the outer layers of the star.   


\begin{figure}
\centering
\includegraphics[scale=0.65]{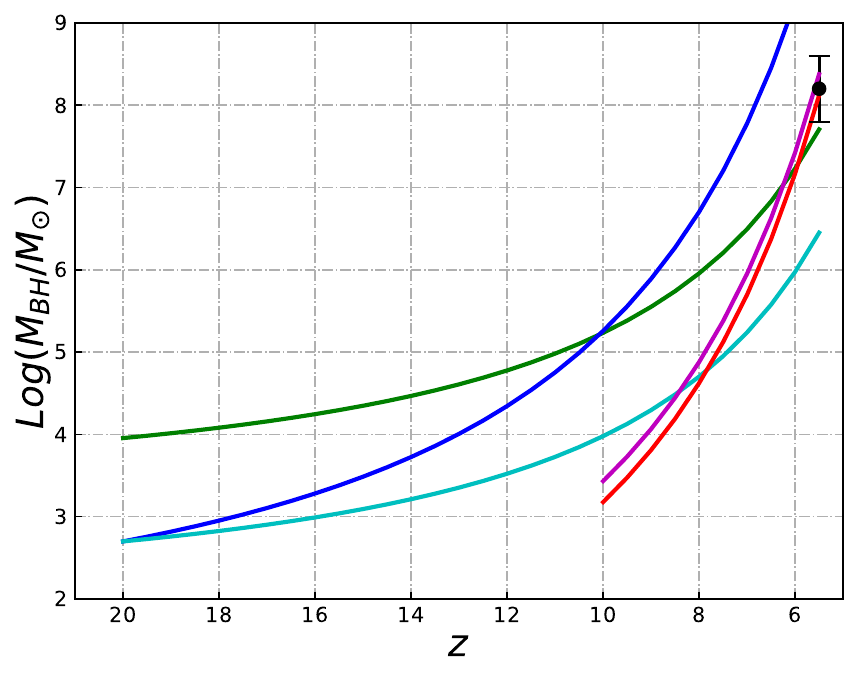} 
\caption{BH growth rates. Birth redshifts are $z =$ 10 and 20 and we consider mass losses of 10\% and 50\%.  The AGN and its error bars in mass are marked by the solid black circle.  Green is the 9000 \Ms\ BH growing at 0.5 $L_{\rm Edd}$, blue is the 500 \Ms\ BH growing at $L_{\rm Edd}$, cyan is the 500 \Ms\ BH growing at 0.5 $L_{\rm Edd}$, magenta is the 2700 \Ms\ BH growing at $L_{\rm Edd}$ and red is the 1500 \Ms\ BH growing at $L_{\rm Edd}$.  Observational constraints on the accretion rate of the AGN at $z =$ 5.55 are 0.1 - 1.6 $L_{\rm Edd}$\cite{ji24}.}
\label{figext:BH}
\end{figure}

Because these stars collapse, massive black holes would be a general feature of high-redshift galaxies with large N/O ratios, and a supermassive star may well have been the origin of the Type-1 AGN in GS 3073, which has log(M$_{\rm BH}$) $=$ 8.2 $\pm$ 0.4 at $z =$ 5.55.  As shown in Extended Data Figure~\ref{figext:BH}, even a 1000 \Ms\ star that has lost 50\% of its mass can reach 10$^{8.2}$ \Ms\ by this redshift at sub-Eddington accretion rates if it is born at $z \gtrsim$ 15.  Such accretion rates are less than those required to form quasars\cite{smidt18,zhu20,latif22b} at $z >$ 6.  Consequently, the discovery of massive BHs in high-redshift galaxies with large N/O ratios in future surveys would reinforce our enrichment scenario.

Because these stars collapse, massive black holes would be a general feature of high-redshift galaxies with large N/O ratios, and a supermassive star may well have been the origin of the Type-1 AGN in GS 3073, which has log(M$_{\rm BH}$) $=$ 8.2 $\pm$ 0.4 at $z =$ 5.55.  Even a 1000 \Ms\ star that has lost 50\% of its mass can reach 10$^{8.2}$ \Ms\ by this redshift at sub-Eddington accretion rates if it is born at $z \gtrsim$ 15.  Such accretion rates are less than those required to form quasars \citep{smidt18,zhu20,latif22b} at $z >$ 6.  Consequently, the discovery of massive BHs in other high-redshift galaxies with large N/O ratios in future surveys would reinforce our enrichment scenario. 

\subsection{Effect of SMS Rotation on Nitrogen Production}


\begin{figure}
\center
\begin{tabular}{c}
\includegraphics[scale=0.25]{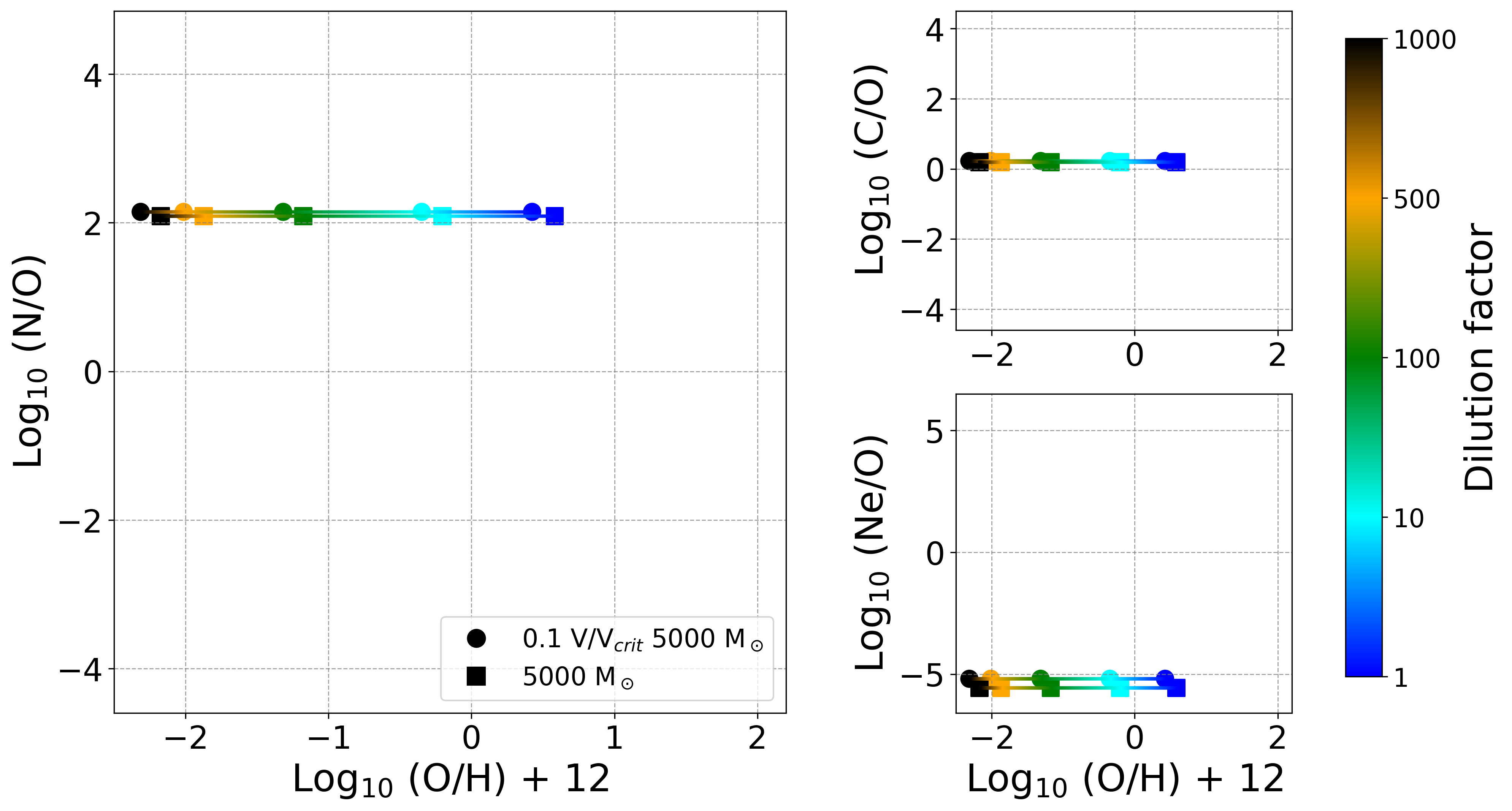}    \\
\includegraphics[scale=0.25]{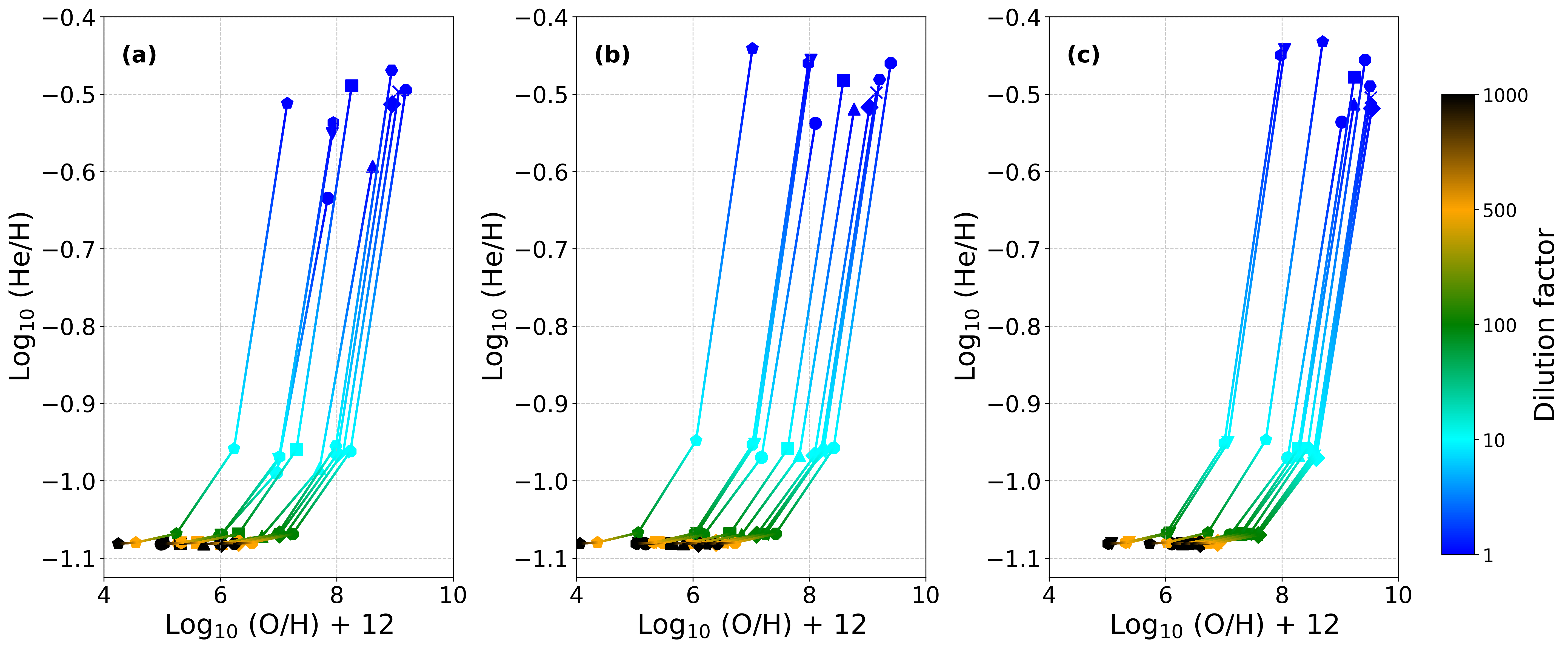} 
\end{tabular}
\caption{\textit{Top}:  N/O ratios for rotating and non-rotating 5000 \Ms\ SMSs.  \textit{Bottom}:  He/H abundance ratios.  Panels (a), (b) and (c) are for mass losses of 10\%, 50\%, and the CO core, respectively.}
\label{fig:rot}
\end{figure}

While rotation is expected to enhance nitrogen production in massive stars, it has never been studied in 1000 - 10,000 \Ms\ stars.  Rotation is challenging to model in Pop III stars above 100 \Ms\ because of the $\Omega$-$\Gamma$ limit, the maximum rotation a star can have without being unbound by radiation and centrifugal forces \citep{Maeder2000}.  Although studies suggest that supermassive stars cannot tolerate much rotation because gravity cannot hold the star together against their extreme outward radiation pressures and centrifugal effects \citep[at most a few percent of Keplerian;][]{hle18a}, they are all expected to have some.  We consider the effect of rotation on nitrogen yields from the 5000 \Ms\ star evolved up to the onset of core helium burning (Y$_c = 0.9$) with an initial rotation rate of 0.1$v/v_{\mathrm{crit}}$. This evolutionary stage is particularly important because the star becomes fully convective at the onset of core helium burning.  N/O ratios with and without rotation are shown for the star at Y$_c = 0.9$ for the case of 10\% mass loss in the upper panel of Figure~\ref{fig:rot}.  We find that rotation results in only a marginal enhancement in N/O, indicating that chemical transport in these stars is primarily dominated by the convective zones that develop during this phase with rotation having a secondary effect.

\subsection{He/H and $^{12}$C/$^{13}$C Ratios}

We show He/H ratios, a prominent observational feature of massive stars \citep{sch03}, for all ten stars and three mass loss cases versus log$_{10}$(O/H) $+$12 measured for GS 3073 in the right panel of Figure~\ref{fig:rot}.  For 10\% mass loss the 5000 and 7000 \Ms\ stars have log$_{10}$(He/H) of -0.55 and -0.53, respectively, at a dilution factor of 1. These ratios are strongly dependent on dilution factor and all the models except 6000 \Ms\ produce log$_{10}$(He/H) from -0.63 to -1.00 at dilution factors of 1 - 200. For 50\% mass loss the 5000 and 7000 \Ms\ stars exhibit log$_{10}$(He/H) of -0.46 and -0.48, respectively, at dilution factors of 1 - 2 while other stars have log$_{10}$(He/H) from -0.65 to 1.00 at dilution factors of 1 - 250. In the case of mass loss down the CO core, the ratios for all the stars are similar to those for 50\% mass loss. He/H ratios are very sensitive to contributions from subsequent populations of stars and change as the galaxy evolves over time. 

$^{12}$C/$^{13}$C ratios are an important signature of CNO burning in massive stars \cite{chi08}.  We find $^{12}$C and $^{13}$C masses to be boosted by mixing between the convective He and H core and shell. For the log$_{10}$(O/H) $+$12 ratio measured for GS 3073, we find log$_{10}$($^{12}$C/$^{13}$C) $=$ 0.59 - 0.65 for dilution factors of 1 - 150. 1000 \Ms\ is an outlier, with log$_{10}$($^{12}$C/$^{13}$C) $=$ 0.800. These ratios are similar for all three cases of mass loss and are consistent with those of rapidly rotating massive stars \citep{chi08}.

\begin{deluxetable*}{cccccccccccc}
\tabletypesize{\scriptsize}            
\setlength{\tabcolsep}{6pt}            
\renewcommand{\arraystretch}{1.00}     
\centerwidetable
\tablecaption{Elemental yields for 1000–10{,}000 \Ms\ Pop III stars for mass losses of 10\%, 50\%, above the CO core, and complete disruption. All masses are in \Ms. $M_{\rm tot}$ is the final mass of the model at the end of evolution, $M_{\rm cont}$ is the mass kept by the star at the time of collapse, and $M_{\rm ej}$ is the mass lost to the ISM.\label{tab:star_ejecta}}
\tablehead{
\colhead{$M_{\rm tot}$} & \colhead{$M_{\rm cont}$} & \colhead{Ejection Criteria} &
\colhead{$M_{\rm ej}$} & \colhead{$M_{\rm H1}$} & \colhead{$M_{\rm He4}$} &
\colhead{$M_{\rm C12}$} & \colhead{$M_{\rm C13}$} & \colhead{$M_{\rm N14}$} &
\colhead{$M_{\rm O16}$} & \colhead{$M_{\rm Ne20}$} & \colhead{$M_{\rm Ne22}$}
}
\startdata
1001.04  & 900.00  & 10\% of $M_{\rm tot}$  & 101.04  & 29.19  & 70.53  & 0.07  & 0.01  & 1.85  & 0.11  & 0.00  & 0.00 \\
2006.89  & 1803.99 & 10\% of $M_{\rm tot}$  & 202.90  & 20.65  & 174.12 & 0.13  & 0.04  & 6.36  & 0.50  & 0.00  & 0.01 \\
3026.37  & 2723.74 & 10\% of $M_{\rm tot}$  & 302.64  & 57.82  & 216.07 & 0.28  & 0.07  & 13.88 & 1.90  & 0.00  & 0.01 \\
3816.99  & 3434.99 & 10\% of $M_{\rm tot}$  & 382.00  & 46.10  & 314.13 & 0.47  & 0.12  & 23.45 & 4.78  & 0.19  & 0.03 \\
5237.93  & 4713.00 & 10\% of $M_{\rm tot}$  & 524.93  & 91.52  & 415.83 & 0.35  & 0.08  & 13.91 & 0.65  & 0.00  & 0.04 \\
6094.62  & 5485.00 & 10\% of $M_{\rm tot}$  & 609.62  & 114.36 & 551.71 & 0.11  & 0.03  & 5.37  & 0.13  & 0.00  & 0.35 \\
7079.48  & 6371.00 & 10\% of $M_{\rm tot}$  & 708.48  & 119.50 & 576.65 & 0.24  & 0.06  & 11.90 & 0.91  & 0.00  & 0.05 \\
8075.33  & 7257.00 & 10\% of $M_{\rm tot}$  & 818.33  & 80.69  & 682.79 & 0.87  & 0.23  & 45.68 & 12.98 & 0.36  & 0.00 \\
9062.10  & 8155.00 & 10\% of $M_{\rm tot}$  & 907.10  & 76.81  & 804.94 & 0.87  & 0.23  & 45.09 & 10.68 & 0.52  & 0.00 \\
10023.87 & 9020.83 & 10\% of $M_{\rm tot}$  & 1003.04 & 104.69 & 843.73 & 0.94  & 0.26  & 44.45 & 20.39 & 4.79  & 0.13 \\
\tableline
1001.04  & 500.54  & 50\% of $M_{\rm tot}$   & 500.50  & 80.50  & 403.04 & 0.72  & 0.11  & 16.07 & 0.92  & 0.00  & 0.00 \\
2006.89  & 1003.89 & 50\% of $M_{\rm tot}$   & 1003.00 & 96.33  & 870.41 & 0.78  & 0.18  & 32.15 & 5.02  & 0.62  & 0.06 \\
3026.37  & 1528.37 & 50\% of $M_{\rm tot}$   & 1498.00 & 191.46 & 1216.13& 1.73  & 0.47  & 93.03 & 12.31 & 0.01  & 0.07 \\
3816.99  & 1908.99 & 50\% of $M_{\rm tot}$   & 1908.00 & 220.40 & 1538.07& 2.45  & 0.59  & 115.46& 27.43 & 2.14  & 0.15 \\
5237.93  & 2619.93 & 50\% of $M_{\rm tot}$   & 2618.00 & 182.80 & 2350.52& 2.08  & 0.56  & 91.53 & 3.67  & 0.00  & 0.51 \\
6094.62  & 3046.62 & 50\% of $M_{\rm tot}$   & 3048.00 & 203.63 & 2845.21& 0.72  & 0.20  & 34.39 & 0.42  & 0.00  & 2.00 \\
7079.48  & 3540.48 & 50\% of $M_{\rm tot}$   & 3539.00 & 274.06 & 3177.47& 1.67  & 0.46  & 84.70 & 4.53  & 0.02  & 0.33 \\
8075.33  & 4038.33 & 50\% of $M_{\rm tot}$   & 4037.00 & 381.23 & 3337.78& 4.82  & 1.15  & 225.00& 74.97 & 5.99  & 0.02 \\
9062.10  & 4531.10 & 50\% of $M_{\rm tot}$   & 4531.00 & 337.23 & 3822.03& 5.62  & 1.13  & 217.05& 93.42 & 19.44 & 0.01 \\
10023.87 & 5012.87 & 50\% of $M_{\rm tot}$   & 5011.00 & 201.49 & 4252.59& 6.32  & 1.75  & 274.42& 158.13& 50.13 & 1.06 \\
\tableline
1001.04  & 298.30  & Above CO core           & 702.74  & 104.29 & 561.61 & 1.64  & 0.15  & 22.88 & 8.54  & 1.40  & 0.00 \\
2006.89  & 482.29  & Above CO core           & 1524.60 & 123.64 & 1319.53& 2.20  & 0.28  & 49.82 & 28.02 & 5.88  & 0.10 \\
3026.37  & 867.43  & Above CO core           & 2158.94 & 227.86 & 1740.31& 4.14  & 0.69  & 138.10& 44.35 & 7.59  & 0.10 \\
3816.99  & 959.29  & Above CO core           & 2857.70 & 262.15 & 2224.26& 7.26  & 0.88  & 173.51& 124.79& 28.99 & 0.23 \\
5237.93  & 1204.70 & Above CO core           & 4033.23 & 202.86 & 3654.76& 3.27  & 0.89  & 145.31& 5.75  & 0.00  & 0.87 \\
6094.62  & 1619.30 & Above CO core           & 4475.32 & 218.68 & 4199.44& 1.85  & 0.30  & 51.95 & 14.84 & 5.25  & 2.99 \\
7079.48  & 2023.50 & Above CO core           & 5055.98 & 315.53 & 4578.28& 2.47  & 0.68  & 125.58& 6.29  & 0.03  & 0.48 \\
8075.33  & 2298.90 & Above CO core           & 5776.43 & 444.48 & 4576.65& 13.39 & 1.63  & 318.27& 244.26& 60.99 & 0.04 \\
9062.10  & 2866.80 & Above CO core           & 6195.30 & 385.41 & 5019.54& 11.72 & 1.52  & 291.23& 248.86& 75.86 & 0.01 \\
10023.87 & 2761.90 & Above CO core           & 7261.97 & 202.45 & 6119.60& 9.67  & 2.67  & 408.68& 242.53& 80.25 & 1.66 \\
\tableline
1001.04  & 0.00    & PISNe                   & 1001.04 & 104.29 & 563.11 & 13.88 & 0.15  & 22.88 & 209.65& 36.51 & 0.00 \\
2006.89  & 0.00    & PISNe                   & 2006.89 & 123.64 & 1322.45& 14.47 & 0.28  & 49.82 & 338.66& 70.60 & 0.11 \\
2484.40  & 0.00    & PISNe                   & 2996.37 & 119.81 & 1336.57& 27.40 & 0.55  & 112.16& 514.50& 125.85& 0.09 \\
3816.99  & 0.00    & PISNe                   & 3816.99 & 262.15 & 2227.61& 27.22 & 0.88  & 173.51& 673.07& 179.25& 0.23 \\
5237.93  & 0.00    & PISNe                   & 5237.93 & 202.90 & 3672.96& 26.13 & 0.89  & 146.07& 593.65& 185.78& 1.03 \\
6094.62  & 0.00    & PISNe                   & 6094.62 & 218.68 & 4199.59& 31.42 & 0.30  & 51.95 & 698.09& 205.56& 3.06 \\
7079.48  & 0.00    & PISNe                   & 7079.48 & 316.26 & 4603.03& 44.28 & 0.69  & 126.30& 1030.16& 289.77& 0.51 \\
8075.33  & 0.00    & PISNe                   & 8075.33 & 444.48 & 4576.80& 55.37 & 1.63  & 318.28& 1308.12& 397.57& 0.09 \\
9062.10  & 0.00    & PISNe                   & 9062.10 & 385.41 & 5019.61& 47.02 & 1.52  & 291.23& 1512.16& 489.28& 0.01 \\
10023.87 & 0.00    & PISNe                   & 10023.87& 202.45 & 6152.83& 13.45 & 2.69  & 411.08& 1213.84& 80.86 & 1.70 \\
\enddata
\end{deluxetable*}

\end{document}